\documentclass[fleqn,12pt,twoside]{article}
\usepackage{espcrc1}
\usepackage{graphicx}
\usepackage[figuresright]{rotating}

\newcommand{\AmS}{{\protect\the\textfont2
  A\kern-.1667em\lower.5ex\hbox{M}\kern-.125emS}}

\title{The Giant Monopole Resonance in the Sn Isotopes: Why is Tin so
``Fluffy''?}

\author{U. Garg\address{Physics Department, University of Notre Dame, \\
Notre Dame, IN 46556, USA}\thanks{Supported in part by the U.S. National
Science Foundation.}}

\begin{document}

\maketitle

\begin{abstract}
The isoscalar giant monopole resonance (GMR) has been investigated in a series
of Sn isotopes (A=112--124) using inelastic scattering of 400-MeV
$\alpha$ particles at extremely forward angles (including 0$^{\circ}$). The
primary aim of the investigation has been to explore the role of the
``symmetry-energy'' term in the expression for nuclear incompressibility. It
is found that the energies of the GMR in the Sn isotopes are significantly
lower than those expected from the nuclear incompressibility previously
extracted from the available data on the compressional-mode giant resonances.
\end{abstract}

\vspace{1cm}
The investigation of the compressional-mode giant resonances---the Isoscalar Giant
Monopole Resonance (GMR) and the Isoscalar Giant Dipole Resonance (ISGDR), an
exotic compressional mode of nuclear oscillation---continues to remain
an active area of work and interest. The primary motivation for the
investigation
of these modes is that they provide a direct experimental determination of the
incompressibility of infinite nuclear matter, K$_{\infty}$, a quantity of
critical importance to understanding the nuclear equation of state.

Experimental identification of these two modes requires inelastic scattering
measurements at extremely-forward angles (including 0$^{\circ}$, where the GMR
Cross sections are maximal). Recent experimental work, using inelastic
scattering of $\alpha$
particles, has been carried out at RCNP, Osaka (400 MeV)
\cite{garg1,garg2,itoh2,uchida,itoh4,itoh3,uchida2,nayak}, at Texas A \& M University
(240 MeV) \cite{dhy,dhy2,henry2,bency,lui,lui2}, and at KVI, Groningen
(200 MeV)
\cite{matyas}. This has been synergistically enhanced by contemporaneous
theoretical work by several groups
\cite{hama,hama2,colo,colo2,colo3,colo4,dario1,dario2,dario3,dario4,jorge1,jorge2,jorge3,jorge4,giai1,giai2,urin,shlomo,shlomo2,shlomo3,abro,misi,kvas,patra};
some of the theory work was previously reviewed by Col\`{o} \cite{colo2} and has
been updated for this Conference \cite{colo4}.

It is now generally accepted that the best method to extract the nuclear
incompressibility, K$_{\infty}$, from the compressional-mode giant resonances,
first proposed by Blaizot \cite{jpb}, is to compare the
experimental GMR energies with the theoretical
values obtained from RPA calculations using different established interactions;
the K$_{\infty}$ associated with the interaction that best reproduces the GMR
(and ISGDR) energies, is considered the ``correct'' experimental value.
Based on this procedure, it has been established
\cite{garg1,uchida,uchida2,colo2,colo3} that both the compressional-mode giant resonances are consistent
with
K$_{\infty} \sim$ 230 MeV. However, when we started the measurements reported
here, the relativistic \cite{dario3,jorge1,giai2} and
non-relativistic calculations \cite{colo2,hama,shlomo} led to significantly
different values for K$_{\infty}$ from the same GMR data, the values from
non-relativistic calculations being, typically, 210--230 MeV and for the
relativistic calculations 250--270 MeV. [Efforts have since been made to create
interactions within the two approaches that are mutually consistent in terms of
the value of the nuclear incompressibility employed
\cite{colo3,jorge4,shlomo3}.] This ``disagreement'' has engendered a healthy debate, and it
appeared that the primary difference between the non-relativistic and
non-relativistic calculations pertained to the values of symmetry energy term
employed; however, the experimental data available prior to the work
reported here did not provide sufficient sensitivity to distinguish between the
two.

The excitation energy of the GMR is expressed in the scaling model \cite{str}
as:
\begin{equation}
E_{GMR}  =  \hbar\sqrt {K_{A}\over m <r^{2}>}
\end{equation}
where K$_{A}$ is the incompressibility of the nucleus and can be expressed as:
\begin{equation}
K_{A} \approx K_{\infty}(1 + cA^{-1/3}) + K_{\tau}((N -Z)/A)^{2} +
K_{Coul}Z^{2}A^{-4/3}.
\end{equation}
Here K$_{\tau}$ and K$_{Coul}$ are negative quantities.
Of these, K$_{Coul}$ is, basically, model independent and the coefficient ``c''
was found to be close to -1 in both relativistic and non-relativistic models.
That leaves K$_{\tau}$ and a more negative
value for this quantity leads to extracting from the experimental K$_{A}$
values a larger value for the K$_{\infty}$ \cite{colo3,jorge2}. 
Indeed, the typical values for
K$_{\tau}$ are $\sim$ -300 MeV and $\sim$ -700 MeV for the ``standard''
non-relativistic and relativistic models, respectively.

The best place to observe a direct effect of this difference in K$_{\tau}$
values is in a series of isotopes where the factor ((N -Z)/A) would vary
significantly {\em without} affecting the other terms in Eq. 2 in any
substantial way. The Sn isotopes afford such an opportunity. Between $^{112}$Sn
and $^{124}$Sn, this factor increases by $\sim$80\% and it was estimated, for
example, that the change in the GMR-energy in going from $^{112}$Sn to $^{124}$Sn would be different in the two calculations by $\sim$0.5 MeV.

We have measured the GMR strength distributions in the Sn isotopes
(A=112,114,116, 118, 120, 122, 124) using inelastic scattering of 400 MeV
$\alpha$ particles. The experiments were carried out at the Research Center
for Nuclear Physics (RCNP), Osaka University.
Details of the experimental procedures have been provided previously
\cite{uchida,itoh4}.
Single-turn-extraction $\alpha$ beams from the Ring Cyclotron
were transported to the Grand Raiden target chamber via a beam-analyzing system
without use of any
slits. Inelastically scattered particles were momentum-analyzed in the
spectrometer. The focal-plane
detector system was comprised of 2 MWDC's and 2 scintillators, providing both
horizontal and vertical positions of incoming particles. The scattering angle at
the target could be determined from ray-tracing. The primary beam was stopped in
three different Faraday cups,
depending on the angle-setting of the spectrometer \cite{itoh4}. Inelastic
scattering data were taken over the angular range 0$^{\circ}$--9$^{\circ}$.
In addition, elastic scattering data were obtained (over the range
3$^{\circ}$--25$^{\circ}$) on all targets to obtain suitable optical-model
parameters for each nucleus under investigation. Calibration data were obtained
using $^{12}$C and $^{24}$Mg targets; a CH$_{2}$ target was employed to account
for the hydrogen-contamination in some of the targets.

A unique aspect of the data obtained in our measurements has been the complete
elimination of
all ``instrumental'' background \cite{garg1,itoh4}. This has been possible
because of the ion-optics of Grand
Raiden: The particles scattered from the target position are focused in the 
vertical direction at
the focal plane while those that undergo a change in trajectory (because of
scattering off the
slits, or the wall or yoke of the spectrometer, for example) are defocused. As a
result, the
``true'' scattering events can be separated from the ``instrumental
background.'' This allows us to carry out further analysis without the need for
arbitrarily subtracting a background from the spectra; the uncertainties
associated with such background-subtraction procedures had
been the bane of giant resonance analyses and a point of constant criticism. In
our analysis, we treat the continuum as composed of a combination of higher
multipoles and include it in the
multipole analysis of the inelastic scattering spectra to extract the
strength-distributions of various multipoles.
The multipole-decomposition procedure is similar to those used previously by
Bonin {\em et al.} \cite{bonin} and by Clark {\em et al.} \cite{henry2}; details
have been provided elsewhere \cite{itoh2,uchida,itoh4,uchida2}.

The ``0$^{\circ}$'' inelastic spectra for the Sn isotopes are presented in
Fig.~1. In all cases, the spectrum is dominated by the GMR peak near E$_{x} \sim$ 15
MeV. The GMR strengths

\vspace*{-1.5cm}
\begin{figure}[!h]
\hspace*{-1.0cm}
\vspace*{-1.2cm}
\includegraphics[height=14.0cm]{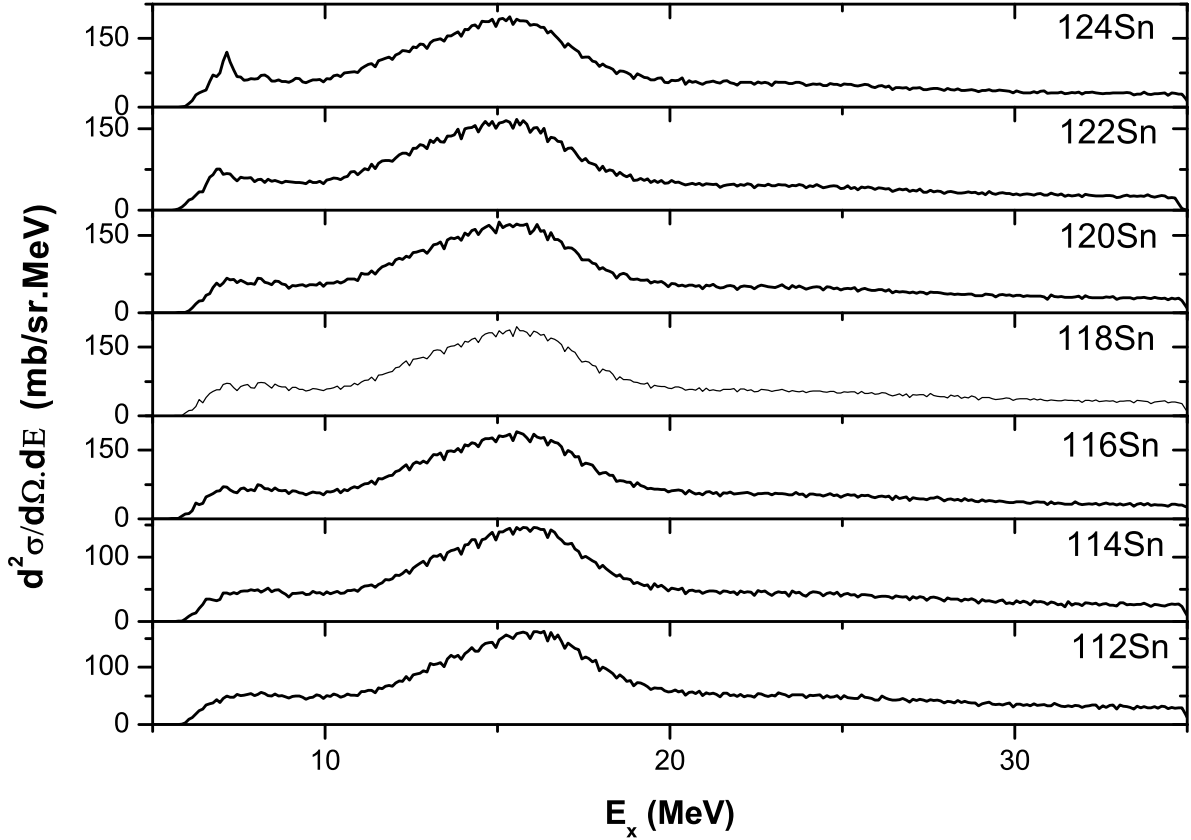}
\vspace*{-1.0cm}
\caption{
Inelastic $\alpha$ scattering spectra at ``0$^{\circ}$'' for the various Sn
targets measured in the present work. The average angle after
accounting for the non-circular shape of the opening of the spectrometer is
0.69$^{\circ}$.
}
\label{spectra.fig}%
\vspace{-0.6cm}
\end{figure}

\noindent 
extracted from the multipole-decomposition analysis are
shown in Fig.~2. The solid lines in Fig.~2 represent Lorentzian fits to the
observed strength distributions. The choice

\begin{figure}[!h]
\begin{center}
\vspace*{-1.2cm}
\hspace*{-0.8cm}
\includegraphics[width=18cm]{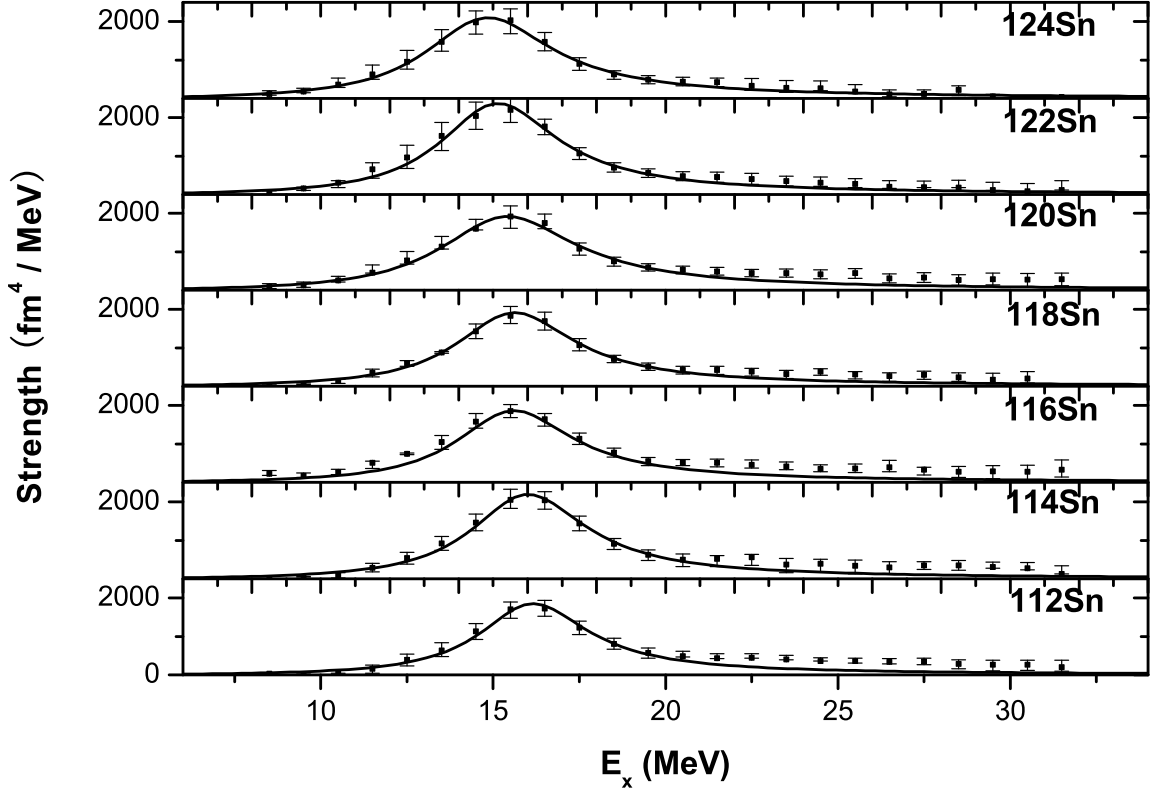}
\vspace*{-1.4cm}
\caption{
Extracted GMR strength distributions in $^{112-124}$Sn from the
multipole-decomposition analysis. The solid lines show results of Lorentzian 
fits to the data.
}
\label{gmr.fig}%
\end{center}
\vspace*{-1.0cm}
\end{figure}

\noindent
of the Lorentzian shape is
arbitrary; the final results are not affected in any significant way by using a
Gaussian shape instead. The extracted GMR parameters, including the various
moment-ratios, are presented in Table 1.

\begin{table} [!h]
\begin{center}
\caption{Lorentzian-fit parameters and various moment-ratios for the GMR
strength distributions in the Sn isotopes, as extracted from
multipole-decomposition analysis in the present work.}
\begin{tabular}{|c|c|c|c|c|c|c|}
\hline
target&$E_0$ (MeV)&$\tau$ (MeV)&$m_1/m_0$ (MeV)& $\sqrt[2]{m_3/m_1}$ (MeV)& $\sqrt[2]{m_1/m_{-1}}$ (MeV) \\
\hline $^{112}$Sn&16.1 $\pm$ 0.09&4.0 $\pm$ 0.42&16.2 $\pm$ 0.13&16.7 $\pm$ 0.15&16.1 $\pm$ 0.12\\
\hline $^{114}$Sn&15.9 $\pm$ 0.14&4.1 $\pm$ 0.38&16.1 $\pm$ 0.12&16.5 $\pm$ 0.17&15.9 $\pm$ 0.11\\
\hline $^{116}$Sn&15.8 $\pm$ 0.13&4.1 $\pm$ 0.33&15.8 $\pm$ 0.10&16.3 $\pm$ 0.16&15.7 $\pm$ 0.12\\
\hline $^{118}$Sn&15.6 $\pm$ 0.08&4.3 $\pm$ 0.38&15.8 $\pm$ 0.11&16.3 $\pm$ 0.14&15.6 $\pm$ 0.13\\
\hline $^{120}$Sn&15.4 $\pm$ 0.19&4.9 $\pm$ 0.54&15.7 $\pm$ 0.10&16.2 $\pm$ 0.15&15.5 $\pm$ 0.11\\
\hline $^{122}$Sn&15.0 $\pm$ 0.16&4.4 $\pm$ 0.41&15.4 $\pm$ 0.10&15.9 $\pm$ 0.18&15.2 $\pm$ 0.11\\
\hline $^{124}$Sn&14.8 $\pm$ 0.21&4.5 $\pm$ 0.52&15.3 $\pm$ 0.10&15.8 $\pm$ 0.14&15.1 $\pm$ 0.11\\
\hline
\end{tabular}
\end{center}
\vspace*{-0.5cm}
\end{table}

Using Eq. (1) and the extracted moment ratios m$_{1}$/m$_{0}$, we have obtained
the values of K$_{A}$ for the Sn isotopes. These are presented in Fig.~3 as a
function of the ``symmetry-parameter'' ((N-Z)/A). A reasonable approximation of
Eq. (2) is that K$_{A}$ has a quadratic
\begin{figure}[!h]
\begin{center}
\vspace*{-1.2cm}
\hspace*{-0.6cm}
\includegraphics[width=18cm]{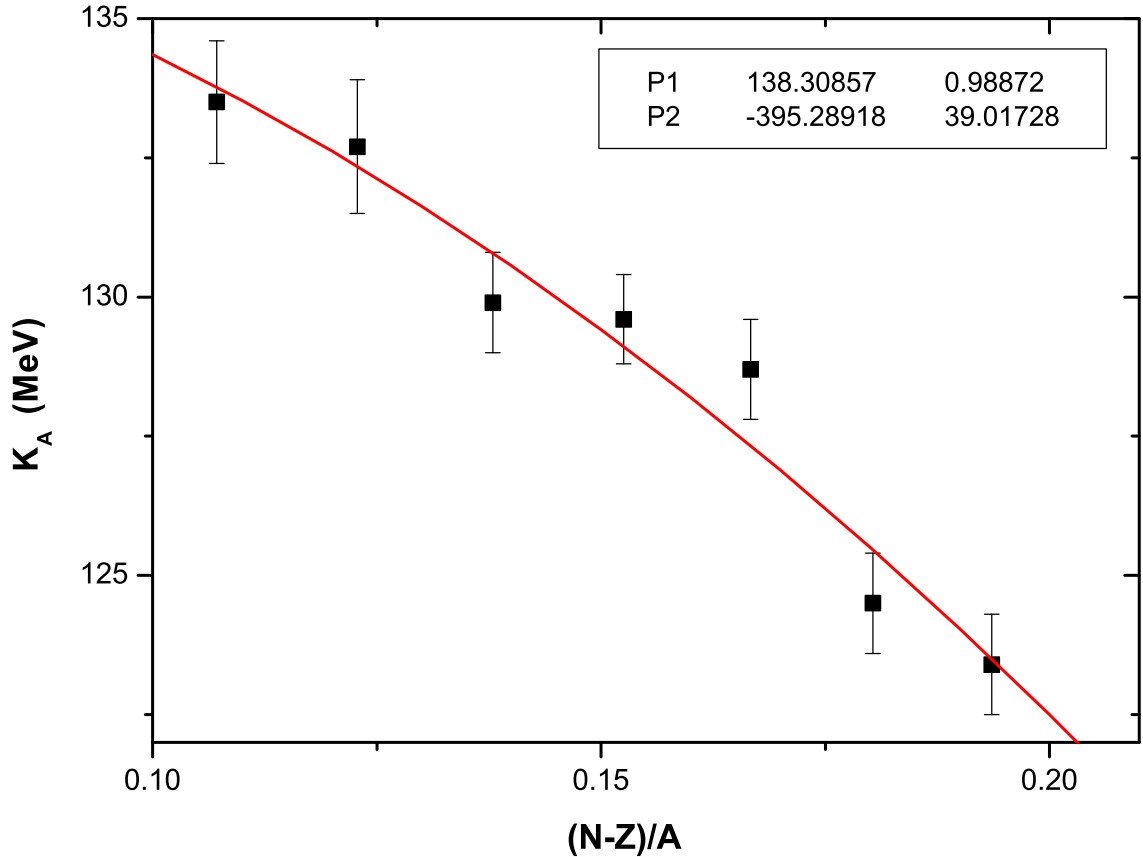}
\vspace*{-1.7cm}
\caption{
Systematics of the values of K$_{A}$ obtained from the moment ratios
m$_{1}$/m$_{0}$ for the GMR strength distributions in the Sn isotopes as a
function of the ``symmetry-parameter'' (N-Z)/A (squares).
A least-squared quadratic fit to the data is shown as a solid line; the
parameters of the fit are shown in the inset.
}
\label{symm.fig}%
\end{center}
\vspace*{-0.7cm}
\end{figure}

\noindent
relationship  with this
``symmetry-parameter'' (of the type C = A + Bx$^{2}$, with the coefficient
``B'' being the parameter K$_{\tau}$). A least-squared quadratic fit to the
data is also shown in the figure. The fit gives a value of
K$_{\tau}$ = -395 $\pm$ 40 MeV. While, admittedly, this value would have a
larger uncertainty if one accounts for the use of the simplified quadratic
equation, and for all possible systematic effects,
it would appear that the symmetry-energy term used in the
non-relativistic calculations ($\sim$ -300 MeV) is closer to the experimental
value than that used in the relativistic calculations ($\sim$ -700 MeV).

The moment-ratios, m$_{1}$/m$_{0}$ for the GMR strengths in the Sn isotopes are
shown in Fig.~4 and compared with recent calculations from Col\`{o}
\cite{colo5} and Piekarewicz \cite{jorgenew}. The interactions used in these
calculations are those that very closely reproduce the GMR energies in
$^{208}$Pb and $^{90}$Zr. But, clearly, the predicted GMR energies for all Sn
isotopes
\begin{figure}[h]
\begin{center}
\vspace*{-0.7cm}
\hspace*{-1.2cm}
\includegraphics[width=18cm]{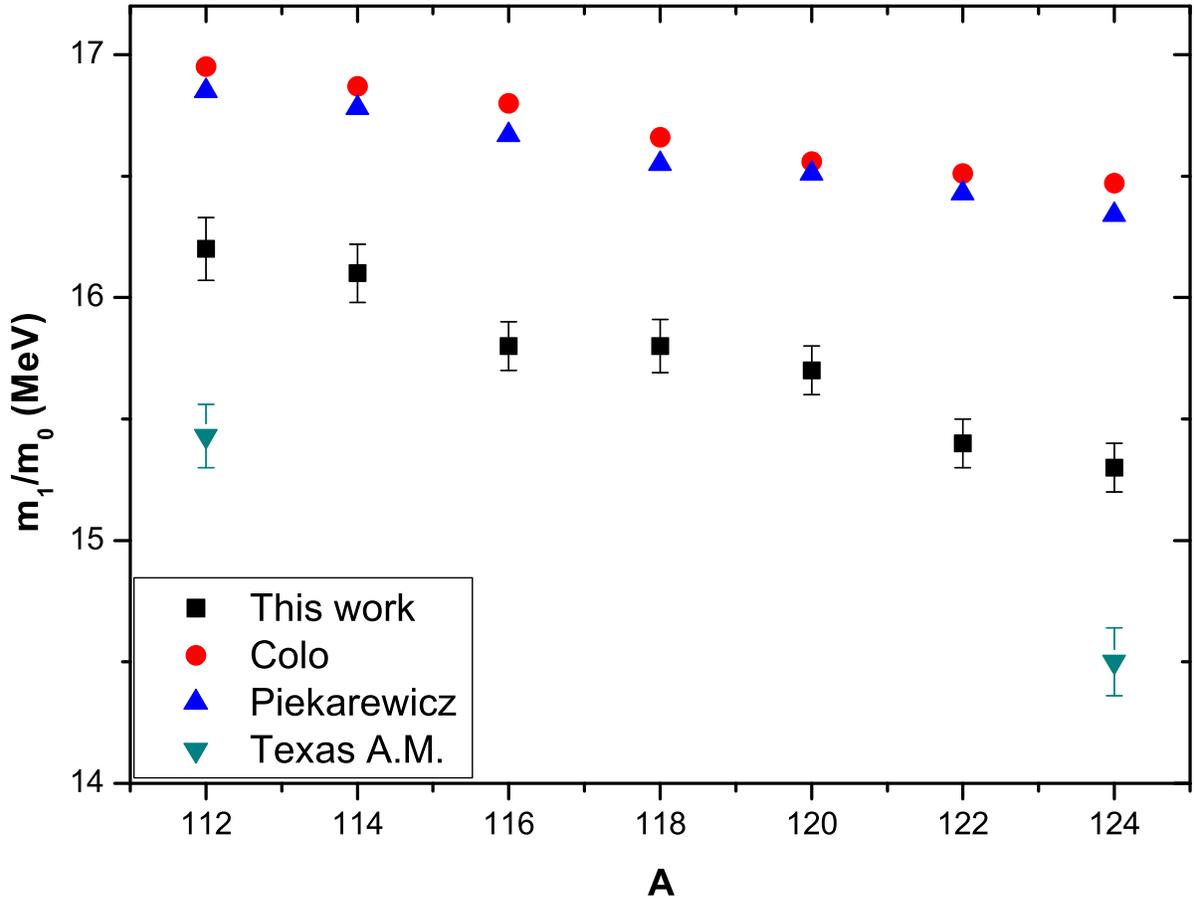}
\vspace*{-1.5cm}
\caption{
Systematics of the moment ratios m$_{1}$/m$_{0}$ for the GMR strength
distributions in the Sn isotopes as a function of mass number (filled squares).
The experimental results are compared with results of very recent calculations
by Col\`{o} \cite{colo5} (filled circles) and Piekarewicz \cite{jorgenew}
(triangles). Also shown are the moment ratios for $^{112}$Sn and $^{124}$Sn,
reported by the Texas A \& M group \cite{lui}(inverse triangles).
}
\label{syst.fig}%
\end{center}
\vspace*{-1cm}
\end{figure}

\noindent
studied in this work are significantly larger than the experimentally
observed values. [Fig.~4 also shows the GMR energies extracted
for $^{112}$Sn and $^{124}$Sn in recent Texas A \& M work; the ``agreement''
with those is even worse!] This leads directly to the question posed in the
title of this report: Why are the Tin isotopes so ``fluffy''? Are there any
nuclear structure effects that need to be taken into account to describe the
GMR energies in the Sn isotopes? Or, more provocatively, do the GMR energies
depend on something more than the nuclear incompressibility, requiring a
modification of the scaling relationship given in Eq. (1)? In the latter case,
why does this ``effect'' show up only in the Sn isotopes? This remains a
challenge to the theoretical calculations describing the GMR.

To summarize, the isoscalar giant monopole resonance (GMR) has been
investigated
in a series of Sn isotopes (A=112--124), using ``small-angle'' inelastic
scattering of 400 MeV particles. The primary aim of these measurements has been
to explore the effect of the ``symmetry-energy'' term, K$_{\tau}$, in the
expression for nuclear incompressibility. The preliminary value of K$_{\tau}$
extracted from these measurements (K$_{\tau}$ = -395 $\pm$ 40 MeV)
is not too different from
the typical values ($\sim$ -300 MeV) employed in the non-relativistic
calculations. It is found that the experimental GMR energies are significantly
lower than those predicted by recent non-relativistic and relativistic
calculations, leaving a challenge for the theories.

It is with pleasure and gratitude that I acknowledge my collaborators in this
work:
T.~Li, Y.~Liu, R.~Marks, B.K.~Nayak, P.V.~Madhusudhana~Rao (Notre Dame);
M.~Fujiwara, H.~Hashimoto, K.~Kawase, K.~Nakanishi, S.~Okumura,
S.~Terashima, M.~Yosoi (RCNP); M.~Itoh (Tohuku Univ.),
T.~Kawabata (Univ. of Tokyo), M.~Uchida (Tokyo Inst. Tech.),
Y.~Iwao, T.~Murakami, H.~Sakaguchi,
Y.~Terashima, Y.~Yasuda, J.~Zenihiro (Kyoto Univ.);
H.~Akimune (Konan Univ.); and, M.N.~Harakeh (KVI).
All of them have worked very hard at the experiments and data analysis, and
the contributions of some of them are undoubtedly greater than my own.
The RCNP ring cyclotron staff merit a special mention for their efforts in
providing high-quality beams for the experiments. This work has been supported 
in part by the National Science Foundation (grants number INT03-42942 and
PHY04-57120).


\begin{thebibliography}{9}
\bibitem{garg1} U. Garg, Nucl. Phys. {\bf A731} (2004) 3.
\bibitem{garg2} M. Hedden {\em et al.}, in {\em Nuclear Physics in the 21st
Century, International Nuclear Physics Conference INPC2001, Berkeley,
California, 2001}, Eric Norman, Lee Schroeder, and Gordon Wozniak, Editors
(American Institute of Physics, Melville, NY, 2002) pp 880.
\bibitem{itoh2} M. Itoh {\em et al.}, Phys. Lett. {\bf B549} (2002) 58.
\bibitem{uchida} M. Uchida {\em et al.}, Phys. Lett. {\bf B557} (2003) 12.
\bibitem{itoh4} M. Itoh {\em et al.}, Phys. Rev. C {\bf 68} (2003) 064602.
\bibitem{itoh3} M. Itoh {\em et al.}, Nucl. Phys. {\bf A731} (2004) 41.
\bibitem{uchida2} M. Uchida {\em et al.}, Phys. Rev. C {\bf 69} (2004)
051301(R).
\bibitem{nayak} B.K. Nayak {\em et al.}, Phys. Lett. {\bf B637} (2006) 43.
\bibitem{dhy} D.H. Youngblood {\em et al.}, Phys. Rev Lett. {\bf 82} (1999)
691.
\bibitem{dhy2} D.H. Youngblood {\em et al.}, Phys. Rev C {\bf 65} (2002)
034302; {\bf 68} (2003) 057303; {\bf 69} (2004) 034315; {\bf 69} (2004)
054312.
\bibitem{henry2} H.L. Clark {\em et al.}, Phys. Rev. C {\bf 63} (2001)
031301(R).
\bibitem{bency} Bency John {\em et al.}, Phys. Rev. C {\bf 68} (2003) 014305.
\bibitem{lui} Y.-W. Lui {\em et al.}, Phys. Rev C {\bf 70} (2004) 014307.
\bibitem{lui2} Y.-W. Lui {\em et al.}, Phys. Rev C {\bf 73} (2006) 014314.
\bibitem{matyas} M. Hunyadi {\em et al.}, Phys. Lett. {\bf B576} (2003) 253.
\bibitem{hama} I. Hamamoto, H. Sagawa, and X.Z. Zhang, Phys. Rev. C {\bf 57}
(1998) R1064.
\bibitem{hama2} I. Hamamoto and H. Sagawa, Phys. Rev. C {\bf 66} (2002) 044315.
\bibitem{colo} G. Col\`{o} {\em et al.}, Phys. Lett. {\bf B485} (2000) 362.
\bibitem{colo2} G. Col\`{o} and Nguyen Van Giai, Nucl. Phys. {\bf A731} (2004)
15.
\bibitem{colo3} G. Col\`{o} {\em et al.}, Phys. Rev. C {\bf 72} (2005) 011302.
\bibitem{colo4} G. Col\`{o}, review talk at this Conference.
\bibitem{dario1} D. Vretenar {\em et al.}, Phys. Rev. C {\bf 65} (2002) 021301.
\bibitem{dario2} D. Vretenar {\em et al.}, Phys. Rev. Lett. {\bf 91} (2003)
262502.
\bibitem{dario3} D. Vretenar, T. Niksic, and P. Ring, Phys. Rev. C {\bf 68}
(2003) 024310.
\bibitem{dario4} G.A. Lalazissis {\em et al.}, Phys. Rev. C {\bf 71} (2005)
024312(R).
\bibitem{jorge1} J. Piekarewicz, Phys. Rev. C {\bf 64} (2001) 024307.
\bibitem{jorge2} J. Piekarewicz, Phys. Rev. C {\bf 66} (2002) 034305.
\bibitem{jorge3} J. Piekarewicz, Phys. Rev. C {\bf 69} (2004) 041301.
\bibitem{jorge4}  B.G. Todd-Rutel and J. Piekarewicz, Phys. Rev. Lett. {\bf 95}
(2005) 122501.
\bibitem{giai1} Nguyen Van Giai and Zhong-yu Ma, RIKEN Review {\bf 23} (1999)
69.
\bibitem{giai2} Zhong-yu Ma {\em et al.}, Nucl. Phys. {\bf A686} (2001) 173.
\bibitem{urin} M.L. Gorelik and M. Urin, Phys. Rev. C {\bf 64} (2001) 047301.
\bibitem{shlomo} S. Shlomo and A.I. Sanzhur, Phys. Rev. C {\bf 65} (2002)
044310.
\bibitem{shlomo2} B.K. Agrawal, S. Shlomo and A.I. Sanzhur, Phys. Rev. C
{\bf 67} (2003) 034314.
\bibitem{shlomo3} B.K. Agrawal, S. Shlomo and V. Kim Au, Phys. Rev. C
{\bf 68} (2003) 031304.
\bibitem{abro} V.I. Abrosimov, A. Dellafiore, and F. Matera, Nucl. Phys.
{\bf A697} (2002) 748.
\bibitem{misi} S. Misicu and S.I. Bastrukov, Eur. Phys. J. {\bf A13} (2002) 399.
\bibitem{kvas} J. Kvasil {\em et al.}, J. Phys. G: Nucl. Part. Phys. {\bf 29}
(2003) 753.
\bibitem{patra} S.K. Patra {\em et al.}, Phys. Rev. C {\bf 65} (2002) 044304.
\bibitem{jpb} J.P. Blaizot {\em et al.}, Nucl. Phys. {\bf A591} (1995) 435.
\bibitem{str} S. Stringari, Phys. Lett. {\bf 108B} (1982) 232.
\bibitem{bonin} B. Bonin {\em et al.}, Nucl. Phys. {\bf A430} (1984) 349.
\bibitem{colo5} G. Col\`{o}, private communication.
\bibitem{jorgenew} J. Piekarewicz, private communication.
\end{thebibliography}
\end{document}